\documentclass[
aps,
prl,
preprint,
groupedaddress,
amsmath,amssymb]{revtex4-1}

\usepackage{graphicx}
\usepackage{dcolumn}
\usepackage{bm}

\begin{document}


\title{Super defocusing of light by optical sub-oscillations}

\author{Yaniv Eliezer}
\email{yaniveli@post.tau.ac.il}
\author{Alon Bahabad}
\affiliation{ Department of Physical Electronics, School of Electrical Engineering, Fleischman Faculty of Engineering, Tel-Aviv University, Tel-Aviv 69978, Israel } 

\date{\today}

\begin{abstract}
	We show that it is possible to construct spectrally lower bound limited functions which can oscillate locally at an arbitrarily low frequency. Such sub-oscillatory functions are complementary to super-oscillatory functions which are band-limited yet can oscillate locally at an arbitrarily high frequency.  We construct a spatially sub-oscillatory optical beam to experimentally demonstrate optical super defocusing. 
\end{abstract}


\maketitle

\section{Introduction}

	In 1988 Aharonov et al \cite{aharonov1988result} have developed the formalism for quantum weak measurements. This formalism relied mathematically on the existence of band-limited signals that  locally oscillate faster than their fastest Fourier components. Such fast oscillations are known today as superoscillations. Super-oscillatory functions were since then explored in many works  \cite{anandan1995quantum,berry2006evolution,berry2008natural,berry2011pointer,berry2013exact,kempf2000black,ferreira2006superoscillations,dennis2008superoscillation,aharonov2011some,katzav2013yield}. The use of superoscillations in quantum weak measurements led to  breakthrough experimental works relating to fundamentals of the quantum measurement process \cite{hosten2008observation,kocsis2011observing,gorodetski2012weak}.
	In 2009, following a suggestion by Berry and Popescu made a few years earlier\cite{berry2006evolution}, a super-oscillatory phenomena was first experimentally demonstrated in optics through the creation of a far field pattern with sub-wavelength hot spots\cite{Huang2009}. This work was followed by a series of works by several groups demonstrating the use of optical super-oscillations in realizing optical super-resolution microscopy \cite{Huang2009, Rogers2012, Wong2013, zheludev2008diffraction,wang2015super}.
	Other successful uses of optical super-oscillations were reported for the demonstration of non-diffracting, accelerating and self-healing super-oscillating optical beams \cite{Greenfield2013,Singh2015,eliezer2016super}, for realizing super-narrow nonlinear frequency conversion \cite{remez2015super} and for light-focusing in the nano-scale regime \cite{david2015nanoscale}. Superoscillations were also suggested as a mean to overcome absorption in dielectric media \cite{EliezerBahabad2014} and for achieving optical temporal super-resolution \cite{Eliezer2016SOB}. Outside of optics, superoscillatory functions were used to realize sub-diffraction focusing of electron beams 
	\cite{remez2016super} and sub-wavelength focusing of radio waves.\cite{wong2010superoscillatory, wong2011sub}.
	
	In this work, we first show theoretically that there exists a complimentary phenomenon to super-oscillations: there exist functions which are lower-bound limited, yet can oscillate at an arbitrarily low rate. We develop this concept of sub-oscillations for functions having either a continuous or a discrete spectrum. Then we use such functions to realize experimentally super defocusing of a light beam.  

\section{Theory} 

	\subsection{Continuous spectrum sub-oscillatory functions} 
	
		M. V. Berry, following Y. Aharonov's argument, has shown \cite{anandan1995quantum} that it is possible to take the following continuous Fourier composition:
		\begin{eqnarray}
			f\left( x, \delta, \alpha \right) = \int\limits_{u_1}^{u_2} {{\rm A}\left( u, \delta, \alpha \right)\exp \left( {ik\left( u \right)x} \right)du} 	
			\label{eq:integralExp}
		\end{eqnarray}			
		
		and make $f\left( x, \delta, \alpha \right)$ oscillate locally faster than it's highest Fourier component (i.e. the highest value of $k(u)$). This is done by using the following distribution function:
		\begin{eqnarray}
			{\rm A}\left( u, \delta, \alpha \right) = \frac{1}{{\sqrt {2\pi } \delta}}\exp \left( { - \frac{{{\left( {u - i\alpha } \right)}^2}}{{2{\delta ^2}}}} \right)
			\label{eq:ampDist}
		\end{eqnarray}	
		
		Which according to Aharonov's reasoning, under the appropriate limit of the  $\delta$ parameter
		acts as a Dirac delta function that shifts the frequency distribution into $k(i\alpha)$ over the complex plane.
		As Berry demonstrated, although $k(u)$ is bound by some upper frequency limit, the complex shifted value $k(i\alpha)$ may exceed this limit and effectively cause the function to super-oscillate, i.e.
		to oscillate locally faster than the fastest frequency component of the band.
		
		For example, if $k(u)$ is selected to be $k_{1}(u) = \cos \left( u \right)$ (the frequency distribution is bound by 1) and $\delta$ is sufficiently small, the complex Dirac delta distribution shifts the frequency distribution into a hyperbolic cosine form:
		\begin{equation}
			\mathop {\lim }\limits_{\delta \rightarrow 0^+} f\left( {x, \delta, \alpha } \right) \cong \exp \left( {ik_{1}\left( {i\alpha } \right)x} \right) = \exp \left( {i\cosh \left( \alpha  \right)x} \right)
		\end{equation}
		Where $k_{1}(i\alpha) = \cosh \left( \alpha  \right)$ with $\alpha > 0$ obviously exceeds the frequency band limit of 1.
		
		Using the same procedure it is possible to take a lower-bound limited frequency distribution signal
		and make it locally sub-oscillate slower than the slowest frequency of its spectrum. 
		This can be done for example by taking a signal with a frequency distribution corresponding to the hyperbolic cosine form $k_{2}(u)=cosh(u)$ which has a lower frequency bound of 1. 
		In this case, the effective frequency distribution resulting due to the shift over the complex plane is the complementary cosine form $k_{2}(i\alpha)=cos(\alpha)$ which is bound from above by 1, which concludes that the signal adopts locally slow frequencies and thus it sub-oscillates.
		
\begin{figure*}[htbp]
			\centering
			\includegraphics[width=1.0\textwidth]{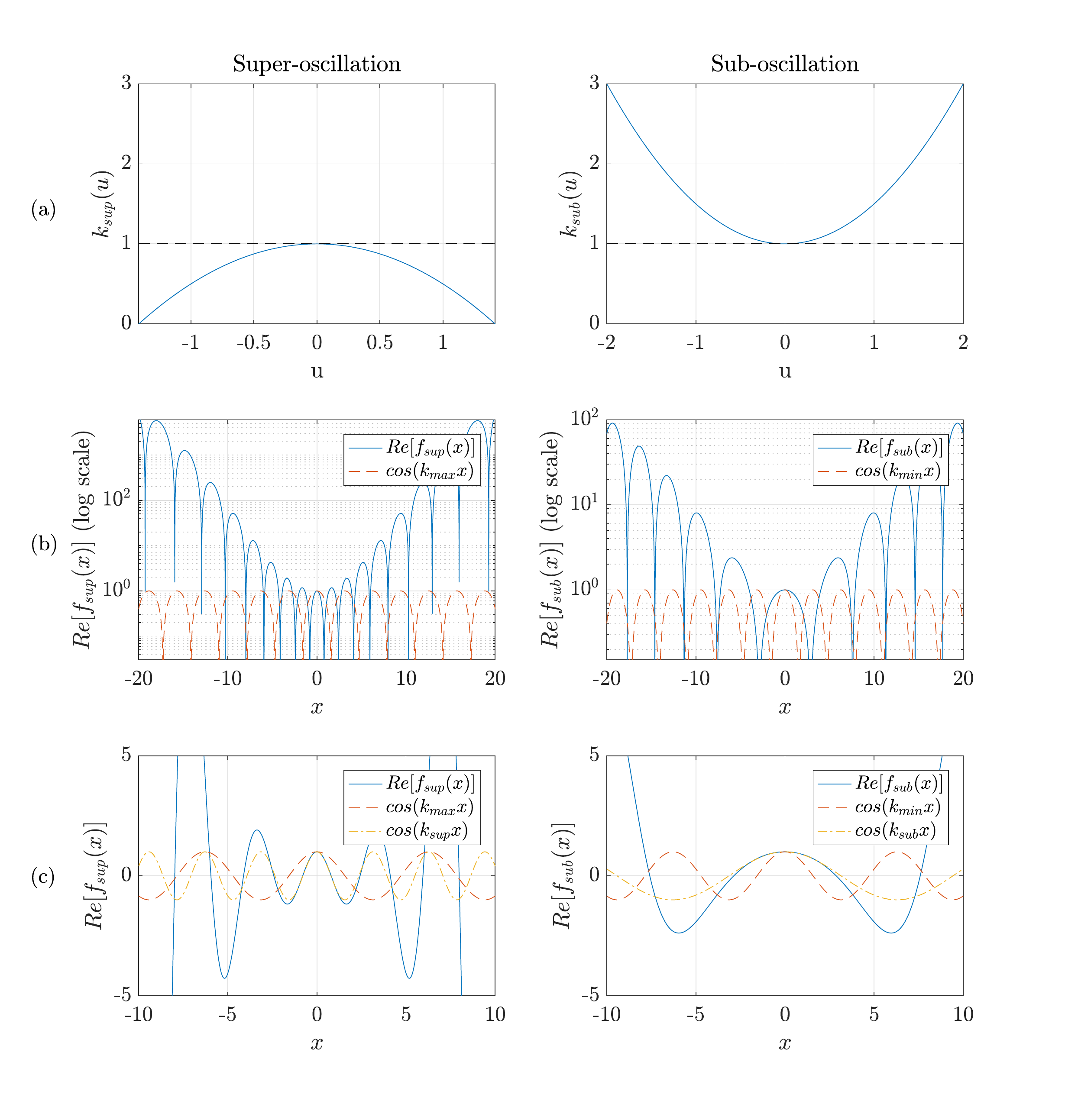}
			\caption{
				Continuous spectrum super-oscillatory (with $\delta = 0.25, \alpha = \sqrt{2}$) and sub-oscillatory ($\delta = 0.25, \alpha = 1$) signals. \textbf{(left)} A super-oscillatory signal. \textbf{(right)} A sub-oscillatory signal. \textbf{(a)} Bound frequency distribution.  \textbf{(b)} Logarithmic scale representation of the real part of the functions (continuous blue line) and logarithmic scale of their highest (lowest) Fourier mode (dashed red line). \textbf{(c)} Linear scale representation  of the signals (continuous blue line), with the most extreme Fourier component (in dashed red line, highest component for the super-oscillatory signal $k_{max}$, lowest component for the sub-oscillatory signal $k_{min}$) and with a Fourier mode not in the spectrum (in dot-dashed yellow line, $k_{sup}$, $k_{sub}$ for the two cases) that matches each function around $x=0$.
			}
			\label{fig1So}
		\end{figure*}

		The integral expression in Eq. \ref{eq:integralExp} fitted with the above frequency distributions $k_{1}(u)$ or $k_{2}(u)$ can be evaluated numerically, while an analytical evaluation is not available at the moment. Nevertheless, Berry's analysis describes a case which can be calculated analytically and can also be approximated using the saddle point  technique. By selecting $k_{3}(u) = 1-\frac{u^2}{2}$ in the range of $-2<u<2$ the function's spectrum is clearly band limited by $[-1,1]$. Applying the complex plane shift operation results in $k_{3}(i\alpha) = 1+\frac{\alpha^2}{2}$, which turns the effective frequency distribution values to larger than 1. 
		The evaluation of Eq. \ref{eq:integralExp} shows the function super-oscillates locally faster than 1 around $x=0$ (see Fig. \ref{fig1So}, left column). 
		
		We use the same methodology to show a sub-oscillatory behavior for the following spectral distribution: 
		$k_{4}(u) = 1+\frac{u^2}{2}$. It is evident that in the same frequency range $-2<u<2$,
		the spectrum's shape is a band pass spectrum exceeding the value of 1.
		Equipped with $k_{4}(u)$ the resulting signal can now be expressed as:
		\begin{equation}
			f\left( {x,\delta ,\alpha } \right) = \frac{1}{{\sqrt {2\pi } \delta }}\int\limits_{ - 2}^2 {\exp \left( {i\left( {1 + \frac{u^2}{2}} \right)x} \right)\exp \left( { - \frac{{{\left( {u - i\alpha } \right)}^2}}{{2{\delta ^2}}}} \right)} du
			\label{eq:cio}
		\end{equation}
		
		which can be evaluated analytically using error functions:
		\begin{eqnarray}
			f\left( {x, \delta ,\alpha } \right) = \frac{i}{{2\sqrt {1 - i\xi} }}\exp \left( {\frac{{ix\left( {2 - {\alpha ^2} - 2i\xi} \right)}}{{2\left( {1 - i\xi} \right)}}} \right) \times \nonumber \\
			\left[ {{\rm{erf}}\left( {\frac{{2 + i\alpha  - 2i\xi}}{{\delta \sqrt {2 - 2i\xi} }}} \right) + {\rm{erf}}\left( {\frac{{2 - i\alpha  - 2i\xi}}{{\delta \sqrt {2 - 2i\xi} }}} \right)} \right]
		\end{eqnarray}			
		where $\xi=x\delta^2$.
		Using the saddle point approximation it is possible to approximate Eq. \ref{eq:cio} with:
		\begin{equation}
			f\left( {x,\delta ,\alpha } \right) \cong \frac{1}{{\sqrt {1 + i\xi} }} 
			\exp \left( ix \left[ 1-\frac{\alpha^2}{2\left(1+\xi^2\right)} \right] \right)
			\exp \left( \frac{\alpha^2 \xi x}{2\left(1+\xi^2\right)} \right)
			\label{isadp}
		\end{equation}	
		
		By differentiating the phase term in the above equation, the following local wave number function is derived:
		\begin{equation}
			q(\xi) = 1 - \frac{\alpha^2(1-\xi^2)}{2(1+\xi^2)^2}
			\label{iclwn}
		\end{equation}		
		
		It is clear that for $|\alpha|<2$ the local frequency around $x=0$ is smaller in its magnitude than  $1$: The signal becomes sub-oscillatory. 
		
		A numerical demonstration of these analytical super-oscillatory and sub-oscillatory functions (i.e. Eq. \ref{eq:integralExp}-\ref{eq:ampDist} for the spectral distribution given with $k_3=k_{sup}$ and $k_4=k_{sub}$ respectively) is shown in 
		Fig. \ref{fig1So}. 
		The logarithmic representation of the functions (Fig. \ref{fig1So}.b) clearly shows that the oscillations of the super-oscillatory (sub-oscillatory) signal accelerate (decelerate) towards $x=0$, while around $x=0$ (Fig. \ref{fig1So}.c) the super-oscillatory (sub-oscillatory) function exhibit an oscillation which is faster (slower) than the highest (slowest) Fourier component of the signal. 
		
	\subsection{Discrete spectrum sub-oscillatory functions} 
	
		Consider the following discrete spectrum super-oscillatory function \cite{aharonov1988result,berry2006evolution}:
		\begin{equation}
			f\left( x \right) = \left(f_1(x)\right)^N = {\left( {\cos (x) + ia\sin \left( x \right)} \right)^N}
			\label{eq:fsodis}
		\end{equation}
		where $a \in \{\mathbb{R}>1\}$ and $N \in \mathbb{N}$.
		This function has the following polar form:
		\begin{eqnarray}
			f\left( x \right) = {\left( {{{\cos }^2}\left( x \right) + {a^2}\sin \left( x \right)} \right)^{N/2}}\exp \left( {iN\arctan \left( {a\tan \left( x \right)} \right)} \right)
			\label{eq:fsopolar}
		\end{eqnarray}
		
		and it can also be expressed using the following band limited Fourier expansion:
		\begin{eqnarray}
			f\left( x \right) = {\left( {\frac{{a + 1}}{2}} \right)^N}\sum\limits_{m = 0}^N {\frac{{{{\left( { - 1} \right)}^m}N!}}{{m!\left( {N - m} \right)!}}} {\left( {\frac{{a - 1}}{{a + 1}}} \right)^m}e^{ {i\left( {N - 2m} \right)x} }
			\label{fsofourier}
		\end{eqnarray}	
		which shows a clear frequency band $[-N,N]$.
		
		Approximating the function $f(x)$ around $x=0$ shows that it locally super-oscillates with the local frequency $N \cdot a$. Calculation of the local frequency function of the super-oscillatory function $f(x)$ shows the same result:
		\begin{equation}
			{k_f}\left( x \right) = {\mathop{\rm Im}\nolimits} \frac{d}{{dx}}\log \left( {f\left( x \right)} \right) =  
			\frac{{Na}}{{{{\cos }^2}x + {a^2}{{\sin }^2}x}}
			\label{fsolocalfreq}
		\end{equation}
		Furthermore, $k_f(x)$ also shows that the function's slowest local frequency is $\frac{N}{a}$ for $x=\frac{\pi}{2} + q\pi \quad$  (where $q \in \mathbb Z$ ). As the signal's spectrum is within the band $[-N,N]$, a selection of an $a$ parameter in the range $0<a<1$ would not result in a slower local oscillation out of the band. To achieve a sub-oscillation, a critical condition is that the function's spectrum should first have a lower-limit. 
		We now show that reciprocating the band-limited,  super-oscillating function of Eq. \ref{eq:fsopolar} results in a function which has a lower-limited spectrum. Then we show that this function also sub-oscillates.
		Reciprocating Eq. \ref{eq:fsopolar} results in:
		\begin{eqnarray}
			g\left( x \right) = {\left( {{{\cos }^2}\left( x \right) + {a^2}\sin \left( x \right)} \right)^{ - N/2}}\exp \left( { - iN\arctan \left( {a\tan \left( x \right)} \right)} \right)
			\label{eq:gio1}
		\end{eqnarray}			
		
		To assure the function in Eq. \ref{eq:gio1} has a lower-limited spectrum, we first reciprocate the base $f_1(x)$ of $f(x)$:
		\begin{equation}
			{g_1}\left( x \right) = {f_1}^{ - 1}\left( x \right) = \frac{1}{{\cos (x) + ia\sin \left( x \right)}}
			\label{g1io}
		\end{equation}			
		
		This function can be expressed as the following Fourier series:
		\begin{equation}
			{g_1}\left( x \right) = \sum_{m=-\infty}^{m=+\infty}{{G^{(1)}_m}\exp(imx)}
			\label{g1iofourier}				
		\end{equation}
		
		and its coefficients can be calculated by complex integration:
		\begin{equation}
			{G^{(1)}_m} = \frac{1}{2\pi}\int\limits_{ - \pi }^\pi  {\frac{{\exp \left( { - imx} \right)dx}}{{\cos (x) + ia\sin \left( x \right)}}}  = \left\{ {\begin{array}{*{20}{c}}
					{2\left[ {\frac{{{{\left( {a + 1} \right)}^{  \frac{1}{2}\left( {m - 1} \right)}}}}{{{{\left( {a - 1} \right)}^{  \frac{1}{2}\left( {m + 1} \right)}}}}} \right],}&{m \in \mathbb{O}^- }\\
					0&{else}
				\end{array}} \right\}
				\label{eq:g1iocoeff}
			\end{equation}
			where $\mathbb{O}^-$ stands for all odd numbers smaller than zero.
			
			The complete Fourier expansion of $g(x)$ can be calculated from Eq. \ref{eq:g1iocoeff} using the following multinomial sum:
			\begin{equation}
				g\left( x \right) = \left({g_1}\left( x \right)\right)^N =  \sum\limits_{\left\{ {\sum {{k_i} = N} } \right\}}^{} {\frac{{N!}}{{{k_1}!{k_2}!...{k_m}!}}} {\left( {G_1^{\left( 1 \right)}{e^{ - ix}}} \right)^{{k_1}}}...{\left( {G_m^{\left( 1 \right)}{e^{ - imx}}} \right)^{{k_m}}}
				\label{eq:giofourier}				
			\end{equation}
			Where the sum is taken over all combinations of the non-negative integer indices $k_1...k_m$ such that $\sum_{i=1}^m{k_i}=N$. 
			
			Clearly, since the spectrum of $g_{1}(x)$ is one sided and the lowest non zero Fourier mode is $m=1$ the N'th power of this Fourier series leads to an infinite number of Fourier modes with the smallest one having frequency of $N$:
			\begin{equation}
				g(x) = \left({G^{(1)}_1}\right)^N\exp(-iNx) + ...
				\label{giolowest}				
			\end{equation}

			Next, by using the local frequency operator on the reciprocal form $g(x)$ (Eq. \ref{eq:gio1}) it is clear that its local frequency function is simply the negation of the local frequency function of the super-oscillatory form $f(x)$ (Eq. \ref{eq:fsopolar}):
			\begin{equation}
				{k_g}\left( x \right) = {\mathop{\rm Im}\nolimits} \frac{d}{{dx}}\log \left( {g\left( x \right)} \right) =  
				-\frac{{Na}}{{{{\cos }^2}x + {a^2}{{\sin }^2}x}}
				\label{giolocalfreq}
			\end{equation}
			
			${k_g}\left( x \right)$ shows the extrema of the local frequency of $g(x)$ are identical to the ones of the original super-oscillatory function. They are $N \cdot a$ and $\frac{N}{a}$.
			However, considering the fact that the reciprocal function has a lower-limited spectrum and that it's slowest frequency is $N$, then by assigning $0<a<1$, the function would 
			oscillate $a$ times slower than the slowest frequency in the spectrum $N$. That is, the function would sub-oscillate at the rate of $N \cdot a$.  
			
			It is also evident that with $0<a<1$ the coefficients in Eq. \ref{eq:g1iocoeff}, \ref{eq:giofourier} decay exponentially with $m$. This property is shared with the super-oscillatory counterpart where the Fourier transform of the function needs to decrease at a sufficient rate
			depending on the magnitude of the super-oscillation \cite{berry2006evolution,aharonov2011some}.
			This natural decay of the spectrum allows to approximate the sub-oscillatory function by omitting the tail of the Fourier coefficients (depending on the accuracy required). Put otherwise, the sub-oscillatory function can be approximated as a lower-limited and band-limited function.
			
\begin{figure*}[htbp]
				\centering
				\includegraphics[width=1.0\textwidth]{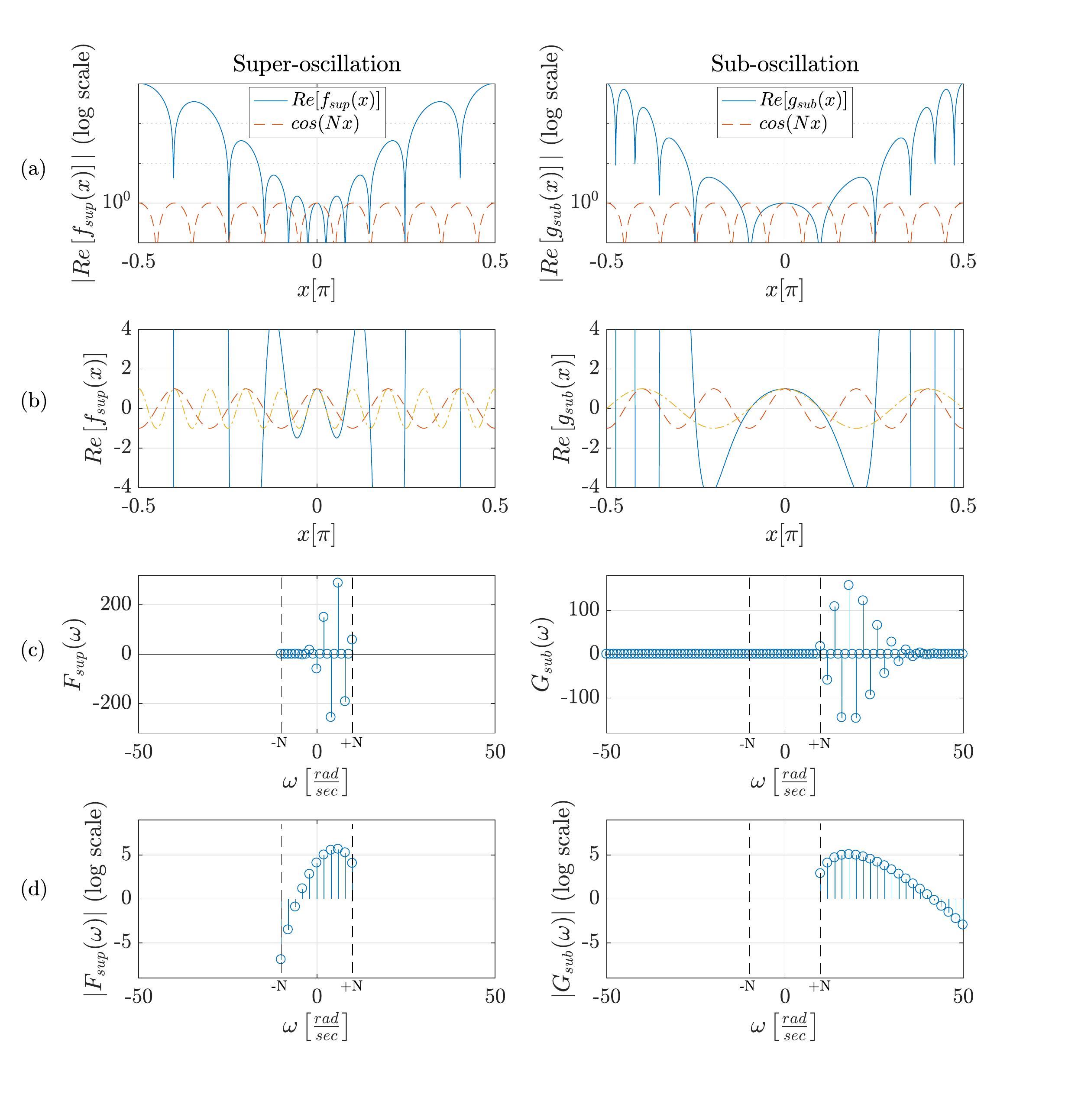}
				\caption{
					Discrete spectrum super-oscillatory and sub-oscillatory signals. \textbf{(left)} Super-oscillatory signal with the parameters of $N=10$ and $a=2$. \textbf{(right)} Sub-oscillatory signal with the parameters of $N=10$ and $a=\frac{1}{2}$.
					\textbf{(a)} Log-scale representations of the functions (continuous blue line) with the most extreme Fourier component (dashed red line, fastest (slowest) for the super-oscillatory (sub-oscillatory) signal). \textbf{(b)} Linear-scale representation of the function (continuous blue line) with the most extreme Fourier component (dashed red line, fastest (slowest) for the super-oscillatory (sub-oscillatory) signal) and with a Fourier mode outside the spectrum (yellow line) which matches the super-oscillation (sub-oscillation). \textbf{(c)} The spectrum in linear scale. The super-oscillatory spectrum is in the range of $\left[-N,N\right]$ while the sub-oscillatory spectrum is in the range of $\left[N,\infty\right]$. \textbf{(d)} The spectrum in logarithmic scale. 
				}
				\label{fig2So}
			\end{figure*}

			A numerical demonstration of the periodic sub-oscillatory function discussed here compared to its reciprocal super-oscillatory function is shown in Fig. \ref{fig2So}. Similarly to the continuous spectrum function discussed in the previous section, the accelerating (deceleration) of the super-oscillatory (sub-oscillatory) function towards $x=0$ is apparent, at which point the function super-oscillate (sub-oscillate) at a rate faster (slower) then the fastest (slowest) component of the spectrum. 
			
	\section{Experiment} 
	
		As an application for sub-oscillatory signals we demonstrate experimentally super-defocusing of a light beam. We consider the following problem: for the strongest defocusing of a light beam we would first focus a given light beam as strongly as possible (the theoretical limit being focusing to a point source) after which the spread in the light beam would be maximal. Now consider the case in which we cannot focus the light beam tightly due to an obstruction by an object in front of the beam but we  still desire to have a strong defocusing after the object. Intuitively we would bring most of the energy of the beam as close as possible to the edges of the obstructing object, to facilitate in the far field a beam with as wide as possible central lobe. Now, equipped with the knowledge of sub-oscillatory functions we can design the field pattern in the plane of the obstructing object with modes extending transversely away from the object such that the central lobe of the beam in the far field would expand arbitrarily fast by making the pattern in the plane of the object in the form of the spectrum of an arbitrarily slow sub-oscillating field.

\begin{figure}[htbp]
			\centerline{\includegraphics[width=1.0\linewidth]{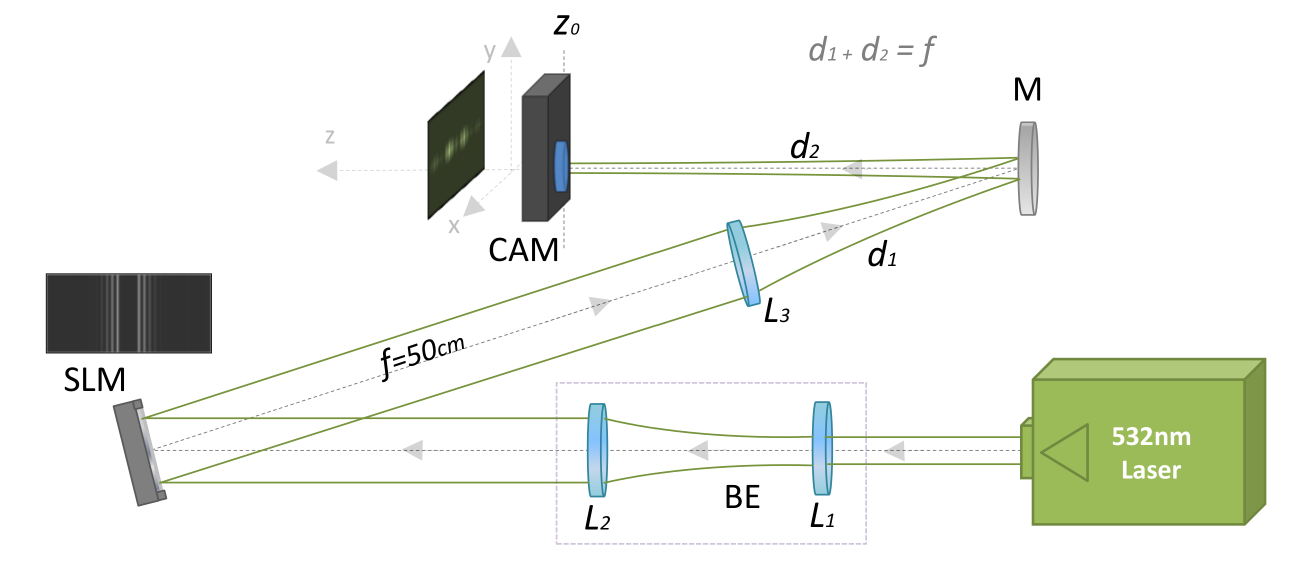}}			
			\caption{ 
				\textbf{Experimental setup.} BE=beam expander. SLM=Spatial Light modulator. CAM=CMOS camera. M=Mirror. $L_1,L_2,L_3$ are lenses. The sum of the distances $d_1,d_2$ equals to lens $L_3$ focal length. $Z_0$ marks the location of the Fourier plane for an object in the SLM plane. 
			}
			\label{fig:ExpSetup}
		\end{figure}

		Our experimental setup (Fig. \ref{fig:ExpSetup}) consists of a 532 nm CW laser (Quantum Ventus 532 Solo Laser) and a reflective phase only Spatial Light Modulator (Holoeye Pluto SLM). The laser light is expanded and collimated before the SLM, reflected of it and Fourier transformed using a 50cm focal lens. The generated beam after the Fourier plane of the lens
		is imaged by a CMOS camera (Ophir Spiricon SP620U Beam Profiling Camera).
		The role of a virtual obstructing object is played by virtue of keeping blank a constant area in the middle of the patterns projected on the SLM 
		
		To demonstrate super defocusing in the experiment we calculated the Fourier transform of the real part of the sub-oscillatory periodic signal in Eq. \ref{eq:gio1} for various values of the parameter $a$ ($a = 1,0.9,0.8,0.7$) with $N=4$. Then we filtered all coefficients having a magnitude which is less than $10^{-14}$ (this corresponds to $M=N+6$ in Eq. \ref{gioexpsum} below) to establish a lower-limited and band limited spectrum. 
		The resulting periodic signal spectra was convoluted with a Gaussian function, to create a corresponding finite Gaussian envelope in the spatial domain. The final shape of the spectrum can be described by the	following expression:
		\begin{eqnarray}
			g\left( k \right) = \sum\limits_{m =  - M}^M {{G_m^{(N)}(a)}\exp\left( { - \frac{{{{\left( {k - {m\Delta}} \right)}^2}}}{{2{\sigma _0}^2}}} \right)} 
			\label{gioexpsum}
			\\
			{G^{(N)}_m}(a) = \frac{1}{\pi}\int\limits_{ - \pi }^\pi
			{\frac{{\cos \left( mx \right)dx}} {{	\left( \cos (x) + ia\sin \left( x \right) \right)^N }}}
			\label{gioexpspect}
		\end{eqnarray}
		
		Where $G_m^{(N)} \equiv 0$ for $|m|<N$, $\Delta = 0.25mm$ and $\sigma_0 = 7.354 \times 10^{-5}$ corresponding to a full width half maximum of $FWHM = 0.173mm$.
		Next we used a first diffraction order phase-only encoding scheme \cite{Bolduc2013} on each of the resulting spectra (that is - for each value of $a$) and projected the phase profile on the SLM. Finally, the field intensity at the back focal Fourier plane (consisting as the far field pattern) was measured using the camera.
		
\begin{figure*}[htbp]
			\hspace*{-2.5cm}
			\includegraphics[width=1.25\textwidth]{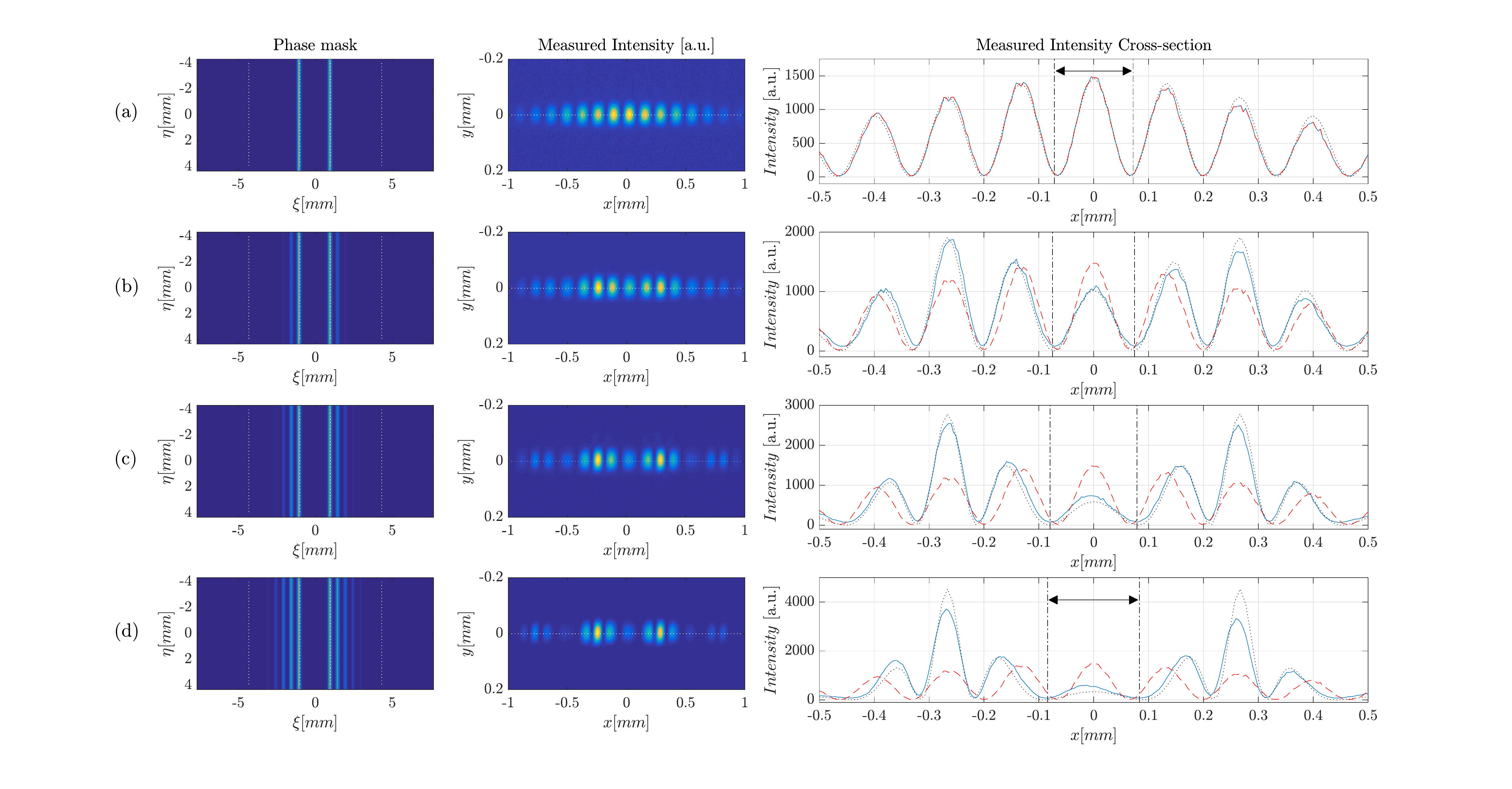}
			\caption{
				Experimental demonstration of super-defocusing of a light beam.
				\textbf{(left)} Phase masks applied to the SLM serving as the spectral distribution for sub-oscillatory signals.
				Dotted white line marks the signal's spectral boundaries.
				\textbf{(middle)} Measured intensity pattern at the plane corresponding to the Fourier plane of the SLM (serving as the far field pattern). \textbf{(right)} Horizontal Line-throughs from the center of the measured intensity patterns (continuous blue lines), theoretically calculated waveforms (dotted black lines) overlaid with the measurement for the regular non sub-oscillatory case (corresponding to $a=1$, dashed red line).
				The different patterns differ by the control parameter $a$. As $a$ gets smaller the pattern sub-oscillates slower and the beam defocuses stronger. \textbf{(a)} a=1 \textbf{(b)} a=0.9 \textbf{(c)} a=0.8 \textbf{(d)} a=0.7. Dotted-dashed black lines mark the location of zeros around the sub-oscillation.
			}
			\label{fig3So}
		\end{figure*}

		The phase masks that were used on the SLM are shown on the left column of Fig. \ref{fig3So}. The masks spatial signature is identical to the spectrum of the sub-oscillatory signals captured on the camera and depicted in the middle column of Fig. \ref{fig3So} (continuous lines). The right column shows an horizontal  line-through from the middle of the captured images. The different rows of this figure correspond from top to bottom to beams which are getting more and more defocused - that is they correspond to signals which are better sub-oscillating. The top row corresponds to a beam which is not sub-oscillatory at all ($a=1$). In this case the corresponding phase mask is composed of only two modes projected into the first diffraction order. The result in the focal (Fourier) plane is a double slit diffraction pattern bound by a Gaussian envelope. The length of each fringe in this case was measured to be $0.1437mm$. For the other rows as $a$ changes between the values of $0.9$ to $0.8$ to $0.7$ (while $N=4$ and the number of Fourier modes is no more than $M=7$) the central lobe of the beam expands correspondingly to $0.1496mm$ to $0.1584mm$ to $0.1672mm$. These three values represents local frequencies which are below the lowest mode of the spectrum which by itself corresponds to the periodicity in the case of $a=1$ (top row of the figure, while the corresponding waveform is overlaid for all cases of different values of $a$ with a dashed line). 
		
		Our results agree reasonably well with calculated theoretical waveforms (dotted lines in Fig. \ref{fig3So}, right column). It is clearly evident that the central sub-oscillation defocuses as the $a$ parameter decreases in a linear rate ($N\cdot a$). It is also apparent that  similarly to super-oscillating signals, the amplitude at the sub-oscillation drops exponentially with respect to its adjacent lobes as the local frequency drops further away from the lower limit of the band. 
		
		
		\section*{Conclusion}
		
		We theoretically and experimentally demonstrated the concept of sub-oscillations, where a spectrally lower-bound limited function oscillates slower than its slowest Fourier component. We have demonstrated theoretically the existence of such functions having either a continuous or a discrete spectrum. Experimentally we used the concept of sub-oscillations to facilitate super-defocusing of a light beam - creating a beam which is initially obstructed in its middle, yet in the far field is having a central lobe which can be arbitrarily wide (albeit at the expanse of its amplitude being smaller relative to the side lobes). 
		
		Sub-oscillations can be regarded as the complementary phenomenon to super-oscillations where a band-limited signal can locally oscillate at an arbitrarily fast rate. The relevance of super-oscillations to varied fields such as quantum measurement \cite{hosten2008observation,kocsis2011observing,gorodetski2012weak}, optical beam shaping and super-resolution \cite{Greenfield2013,Singh2015,eliezer2016super,Huang2009, Rogers2012,Wong2013,zheludev2008diffraction,wang2015super}, particle manipulation \cite{brijesh2016}, electron beam shaping \cite{remez2016super} and radio frequency antenna design \cite{wong2010superoscillatory,wong2011sub},  suggest that sub-oscillations could find interesting uses in varied fields as well. Our demonstration of super defocusing by itself might be relevant for optical  dark-field microscopy. 
		
	
		

	\bibliography{bibliography}

\end{document}